\documentclass[showpacs,preprintnumbers,amsmath,amssymb, 
twocolumn, tightenlines,
]{revtex4}

\usepackage{bm}

\usepackage{amsmath}

\usepackage{dcolumn}

   \usepackage[dvips]{graphicx}
\begin{document}

    \bibliographystyle{apsrev}
    
    \title {Coulomb problem for vector bosons versus Standard Model}
    
    \author{M.Yu.Kuchiev$^1$}
    \email[Email:]{kuchiev@phys.unsw.edu.au}
    \author{V.V.Flambaum$^{1,2}$}
    \email[Email:]{flambaum@phys.unsw.edu.au}
    
    \affiliation{$^1$School of Physics, University of New South Wales,
      Sydney 2052, Australia\\
      $^2$ Physics Division, Argonne National Laboratory, Argonne,
      Illinois 60439-4843, USA }

    \date{\today}

    \begin{abstract} 
      The Coulomb problem for vector bosons $W^\pm$ propagating in an
      attractive Coulomb field incorporates a known difficulty, i.e.
      the total charge of the boson localized on the Coulomb center
      turns out infinite. This fact contradicts the renormalizability
      of the Standard model, which presumes that at small distances
      all physical quantities are well defined. The paradox is shown
      to be resolved by the QED vacuum polarization, which brings in a
      strong effective repulsion and eradicates the infinite charge of
      the boson on the Coulomb center. The effect makes the Coulomb
      problem for vector bosons well defined and consistent with the
      Standard Model.
    \end{abstract}
    

    \maketitle

    Consider a massive charged vector boson $W^{\pm}$ propagating in
    an attractive Coulomb field created by a point-like, heavy
    particle. (A small primordial charged black hole gives an example
    of such particle since its Coulomb interaction with a vector boson
    is much stronger than their gravitational attraction.)  It has
    been ``always'' known that this problem incorporates a difficulty.
    Shortly after Proca formulated his theory for vector particles
    \cite{proca_1936} it became clear that it produces inadequate
    results for the Coulomb problem
    \cite{massey-corben_1939,oppenheimer-snyder-serber_1940,tamm_1940-1-2}.
    This prompted Corben and Schwinger \cite{corben-schwinger_1940} to
    modify the Proca theory by tuning one coefficient in the
    Lagrangian and equations of motion and forcing the $g$-factor of
    the vector boson to take a favorable value $g=2$. It was
    recognized later \cite{schwinger_1964} that the formalism of
    Ref.\cite{corben-schwinger_1940} has a close connection with the
    non-Abelian gauge theory, which makes it relevant for the present
    day studies.  To emphasize this important point we outline below a
    derivation of the Corben- Schwinger equation directly from the
    Standard Model.  The necessity for vector particles to have an
    anomalous magnetic ratio $g=2$ has been thoroughly discussed in
    literature, see e.g.  Ref.\cite{cheng_wu_1972,huang_1992}.
   
    Ref.\cite{corben-schwinger_1940} found that the discrete energy
    spectrum of the Coulomb problem for vector bosons is simple and
    realistic; Ref.\cite{kuchiev_flambaum_2005} re-derived this
    important result. However, Ref.\cite{corben-schwinger_1940} also
    discovered a fundamental flaw in the problem.  For quantum states
    with the total angular momentum zero, $j=0$, the charge density of
    the boson is singular at the origin, which makes the total charge
    divergent there and therefore implies that the Coulomb problem is
    poorly defined.  This is unsatisfactory because the effect
    manifests itself for any, however small value of the Coulomb
    charge $Z$; moreover, it takes place at small distances, where the
    Standard Model, being a renormalizable theory
    Ref.\cite{thooft-veltman-1972}, should not encounter problems of
    this kind.  Thus, there exists a clear contradiction.  The Coulomb
    problem derived from the Standard Model produces results, which
    challenge the Model itself.
    
    This known difficulty has motivated several lines of research.
    Early efforts are summarized in
    Ref.\cite{vijayalakshmi-seetharaman-mathews_1979}.  More recent
    Refs.  \cite{pomeransky-khriplovich_1998,pomeransky-se'nkov_1999,%
      pomeransky-sen'kov-khriplovich_2000} suggested a new, refined
    modification of the formalism for vector bosons.
    Ref.\cite{silenko_2004} claimed that this new formalism complied
    with results of Corben and Schwinger. Some authors considered
    other equations governing vector bosons
    \cite{fushchych-nikitin-susloparow_1985,fushchych-nikitin_1994,%
      sergheyev_1997}, which differ substantially from the
    Corben-Schwinger one, producing more acceptable results for the
    Coulomb problem. However, these approaches could not be based on a
    renormalizable theory.
    
    Overall, in spite of the progress made over the years, the
    contradiction related to the inconsistency of the Coulomb problem
    for vector bosons and renormalizability of the Standard Model
    still exists. We find a clear way to resolve it, i.e. to formulate
    the Coulomb problem for vector particles properly, within the
    framework of the Standard Model. Our main observation is that the
    polarization of the QED vacuum has a profound impact on the
    problem forcing the density of charge of a vector boson to
    decrease exponentially at the origin, thus making the Coulomb
    problem stable and well defined.
    
    At first glance this result looks surprising because the vacuum
    polarization makes the attractive field stronger at the origin
    and, presumably, increasing the charge density at the origin.
    Moreover, the vacuum polarization for spinor and scalar particles
    in the Coulomb field is known to produce only small, perturbative
    effects.  In contrast, for the vector particle we find a {\it
      strong reduction} in the charge density.  To grasp a physical
    mechanism involved it is necessary to notice that the equation of
    motion for vector particles contains a particular term, which has
    no counterparts for scalars and spinors (see the last term in
    Eq.(\ref{form})).  It is this additional term that is responsible
    for a strong effective repulsion, which makes the Coulomb problem
    stable.

    Consider boson fields in the Lagrangian of the Standard Model, see
    e.g. Ref.\cite{weinberg_2001} ($\hbar=c=1$),
   \begin{eqnarray}
      \label{gauge}
      {\cal L}= -\frac{1}{4}\,
      \left(\partial_\mu \boldsymbol{A}_\nu-\partial_\nu \boldsymbol{A}_\mu  +
        g \,\boldsymbol{A}_\mu \times \boldsymbol{A}_\nu\right)^2,
      \\ \nonumber
            -\frac{1}{4}\,
      \left(\partial_\mu { B}_\nu-\partial_\nu {B}_\mu  \right)^2+
      \frac{1}{2}\,D_\mu\Phi^+ D^\mu \Phi~.
    \end{eqnarray}
    Here $\boldsymbol{A}_\mu$ and $B_\mu$ are the $SU(2)$ and $U(1)$
    gauge potentials respectively (abridged notation is used here).
    The covariant derivative $D_\mu\Phi$ takes into account the fact
    that the Higgs field $\Phi$ has a hypercharge, which describes its
    interaction with the $U(1)$ field, and is transformed as a doublet
    under the $SU(2)$ gauge transformations, $\Phi=(\phi_1,\phi_2)$.
    Taking the unitary gauge, one can present $\Phi$ via one real
    component $\Phi=(0,\phi),~\phi=\phi^+$.  Assuming that the scalar
    field develops the vacuum expectation value $\phi=\phi_0$, one
    derives from Eq.(\ref{gauge}) that there appears an
    electromagnetic field $A_\mu =-\sin \theta \,A_\mu^3+\cos\theta
    \,B_\mu$ ($\theta$ is the Weinberg angle) and a pair of charged
    massive vector bosons $W_\mu \equiv W^-_\mu=(A_\mu^1-iA_\mu^2
    )/\sqrt 2$, and $W^+_\mu\equiv(W_\mu)^+$.
    
    Expanding the Lagrangian in the vicinity of $\phi=\phi_0$ and
    retaining only bilinear in $W_\mu,W_\mu^+$ terms for massive
    vector bosons, one derives from Eq.(\ref{gauge}) an effective
    Lagrangian which describes propagation of $W$-bosons in an
    external electromagnetic field
    \begin{eqnarray}
      \nonumber
      {\cal L}^W &=& -\frac{1}{2}\left(\nabla_\mu W_\nu -
        \nabla_\nu W_\mu\right)^+ \left(\nabla^\mu W^\nu-
        \nabla^\nu W^\mu\right)  
      \\ \label{W}      
        &&+ i e \, F^{\mu\nu} \,W^+_\mu W_\nu + 
        m^2 \,W_\mu^+ W^\mu~. 
    \end{eqnarray}
    Here $m$ is the mass of $W$. The external field is accounted for
    in Eq.(\ref{W}) by the derivative $ \nabla_\mu=\partial_\mu +i e
    A_\mu$, and by the term with the field $F^{\mu\nu}=\partial^\mu
    A^\nu-\partial^\nu A^\mu$, which was first introduced into the
    Lagrangian for vector bosons in \cite{corben-schwinger_1940}.
    From Eq.(\ref{W}) one derives the classical Lagrange equation of
    motion for vector bosons
    \begin{eqnarray}
      \label{wave}
      \left( \nabla^2+m^2\right) W^\mu
      + 2 i e \,F^{\mu\nu}\,W_\nu-  
      \nabla^\mu \nabla^\nu \,W_\nu =0~.
    \end{eqnarray}
    Here the identity $[\nabla_\mu,\nabla_\nu]=ieF_{\mu\nu}$ was used.
    Taking a covariant derivative in Eq.(\ref{wave}) one finds
    \begin{eqnarray}
      \label{Lgauge}
      m^2\nabla_\mu W^\mu+ie\,j_\mu W^\mu=0~,
    \end{eqnarray}
    where $j^\mu=\partial_\nu F^{\nu\mu}$ is the external current.
    Substituting $\nabla_\mu W^\mu$ from Eq.(\ref{Lgauge}) back into
    Eq.(\ref{wave}) one can rewrite the latter in a more transparent
    form
    \begin{eqnarray}
      \label{form}
      \left( \nabla^2+m^2\right) W^\mu
      + 2 i e \,F^{\mu\nu}\,W_\nu+ \frac{ie}{m^2}  
      \nabla^\mu (j_\nu W^\nu) =0.
    \end{eqnarray}
    We will use below the current of vector bosons
    $j_{\phantom{\,}\mu}^{W}$, which is obtained by taking a variation
    of the Lagrangian Eq.(\ref{W}) with respect to $A_\mu$
    \begin{eqnarray}
      \label{jw}
      j_{\phantom{\,}\mu}^{W}=-ie\Big( \,W^+_\nu \nabla_\mu W^\nu + 
      2 (\nabla_\nu W^+_\mu)  W^\nu -c.c. \,\Big)
      \\ \nonumber 
     -\frac{e^2}{m^2}(\, W_\mu^{+} W_\nu+W_\nu^{+}W_\mu\,)\,j^\nu~.
    \end{eqnarray}
    Here $c.c.$ refers to two complex conjugated terms and
    Eq.(\ref{Lgauge}) was employed to present the current in this
    form.
      
    Consider now the case of a static electric field described by the
    electric potential $A_0=A_0(\boldsymbol{r})$ and charge density
    $\rho=\rho(\boldsymbol{r})=-\Delta A_0$. For a stationary state of
    the $W$-boson one can presume that $\nabla_0 W_\mu=
    -i(\varepsilon-U)W_\mu$, where $\varepsilon$ is the energy of the
    state, and $U=U(\boldsymbol{r})= eA_0$ is the potential energy of
    the $W$-boson in the electric field.  Eq.(\ref{Lgauge}) in this
    case gives
    \begin{eqnarray}
      \label{simp}
      (\varepsilon-U-\Upsilon)\,w= \boldsymbol{ \nabla \cdot W}~.
    \end{eqnarray}
    The four-vector $W^\mu=(W_0,\boldsymbol{W})$ is presented here via
    the three-vector $\boldsymbol{W}$ and a convenient parameter $w= i
    W_0$.  In order to simplify notation we introduce also an
    important quantity $\Upsilon=\Upsilon(\boldsymbol{r})$,
    \begin{eqnarray}
      \label{Ups}
      \Upsilon=e\rho/m^2=-\Delta U/m^2~.
    \end{eqnarray}
    Eq.(\ref{form}) can be conveniently presented in this notation
    \begin{eqnarray}
      \label{wa}
       \big((\varepsilon-U)^2-m^2\big)\boldsymbol{W}=-\Delta
      \boldsymbol{W}-2\boldsymbol{\nabla} U w-\boldsymbol{\nabla}(\Upsilon w)~. 
    \end{eqnarray}
    For a spherically symmetrical case $U=U(r),~\Upsilon=\Upsilon(r)$
    the total angular momentum $j$ conserves.  We restrict our
    discussion below to the most important case of longitudinal states
    that have $j=0$, where the field $\boldsymbol{W}$ can be presented
    with the help of a radial function $v=v(r)$
      \begin{eqnarray}
        \label{v}
      \boldsymbol{ W } =v \,\boldsymbol{n},\quad \quad 
      \boldsymbol{n}= \boldsymbol{r}/r ~.
      \end{eqnarray}
      Eq.(\ref{simp}) allows one to express the function $w$ via $v$
      \begin{eqnarray}
        \label{wv}
      w= \left(\varepsilon-U-\Upsilon\right)^{-1}\,\big(v'+2v/r\big)~.
      \end{eqnarray}
      Substituting $\boldsymbol{W}$ and $w$ from
      Eqs.(\ref{v}),(\ref{wv}) into Eq.(\ref{wa}) one finds an
      equation on $v$
      \begin{eqnarray}
        \label{sec}
        v''+G \, v'+H \,v=0~,
      \end{eqnarray}
      where the coefficients $G=G(r)$ and $H=H(r)$ are
      \begin{eqnarray}
        \label{g}
        &&G = \frac{2}{r}+\frac{U'}{\varepsilon-U}+\frac{U'+\Upsilon'}
        {\varepsilon-U-\Upsilon}~,
        \\ \label{h}
        &&H =-\frac{2}{r^2}+\frac{2}{r}
        \left( \frac{U'}{\varepsilon-U}+\frac{U'+\Upsilon'}{\varepsilon-U-\Upsilon}
        \right)
        \\ \nonumber
        &&\quad \quad \quad +\frac{\varepsilon-U-\Upsilon}{\varepsilon-U}
        \big( \, (\varepsilon-U)^2-m^2 \,\big)~.
      \end{eqnarray}
      It is convenient to scale the radial function
      $v\rightarrow\varphi=\varphi(r)$
      \begin{eqnarray}
        \label{phi}
        v=\varphi\big[ \, \big(\varepsilon-U(r)\,\big)
        \big(\varepsilon-U(r)-\Upsilon(r)\,\big)
        \,\big]^{1/2}/r.
      \end{eqnarray}
      Then one writes the equation of motion for $W$-bosons
      Eq.(\ref{sec}) in an appealing form
      \begin{eqnarray}
        \label{phi''}
       && -\varphi''+{\cal U}\,\varphi=0~.
        \\ \label{ugh}
        &&{\cal U}=-\,H+G^{\,2}/4+G{\,'}/2~,
      \end{eqnarray}
      $G,H$ are defined in Eqs.(\ref{g}),(\ref{h}).  Eq.(\ref{phi''})
      can be looked at as a Schr\"odinger-type equation, in which
      ${\cal U}={\cal U}(r)$ plays the role of an effective potential
      energy.
      
      We will need below an expression for the charge density
      $\rho^{W}=j_{\phantom{\,}0}^{W}$ of a vector boson in the $j=0$
      state. From Eqs.(\ref{jw}),(\ref{v}) one derives
      \begin{eqnarray}
        \label{expr}
       \rho^W=2 e\big[\,(\varepsilon-U)(v^2+w^2)+2vw'-\Upsilon w^2\big]~.
      \end{eqnarray}
      Let us study properties of solutions of Eq.(\ref{phi''}).
      Consider first the simplest example, the pure Coulomb potential
      \begin{eqnarray}
        \label{coul}
      U(r)=-Z\alpha/r~,\quad\Upsilon(r)=0~,\quad r>0~.
      \end{eqnarray}
      Then Eq.(\ref{ugh}) shows that for small distances $r\ll
      Z\alpha/m$ the effective potential satisfies
      \begin{eqnarray}
        \label{sati}
        {\cal U}(r) \simeq -U^2(r)=-(Z\alpha)^2/r^2~. 
      \end{eqnarray}
      Consequently, a solution of Eq.(\ref{phi}) regular at $r=0$
      exhibits the behavior
      \begin{eqnarray}
        \label{beha}
        \varphi(r) \propto r^{\gamma+1/2}~,  \quad
        r\rightarrow 0~.
      \end{eqnarray}
      Here $\gamma =(\,1/4-(Z\alpha)^2 \,)^{1/2}$.  At first sight
      Eq.(\ref{beha}) looks harmless, but in fact it leads to a
      fundamental problem with the charge density.
      Eqs.(\ref{wv}),(\ref{phi}),(\ref{beha}) give $v\propto Z\alpha
      r^{\gamma-3/2}$, $w\propto(\gamma+1/2) r^{\gamma-3/2}$.  From
      Eq.(\ref{expr}) one estimates the charge density,
      $\rho^{W}\propto r^{2\gamma-4}$, finding it so singular at the
      origin that the total charge $\int \rho^{W}d^3r$ is divergent
      there. This important result was discovered by Corben and
      Schwinger \cite{corben-schwinger_1940}.  One has to conclude
      that the pure Coulomb problem for vector bosons in $j=0$ state
      is poorly defined.
      
      Consider the conventional QED vacuum polarization. The potential
      energy in this case can be written as
      \begin{eqnarray}
        \label{pot}
        U(r)=-\big(\,1+S(r)\,\big) Z\alpha/r~,
      \end{eqnarray}
      where $S(r)$ accounts for the polarization.  It
      suffices to consider the effect in the lowest-order
      approximation, when the polarization is described by the Uehling
      potential. Its small-distance expansion, see e.g. \cite{LL4},
      reads
      \begin{eqnarray}
        \label{write}
        S(r)\simeq -\alpha \beta \,\ln\left(m_{Z}r\right),
        \quad
        r\rightarrow 0.
      \end{eqnarray}
      The logarithm here is closely related to the logarithm
      responsible for the scaling of the QED coupling constant
      $\alpha^{-1}(\mu)=\alpha^{-1}(\mu_0)-\beta\ln(\mu/\mu_0)$, see
      e.g. \cite{LL4}. The factor $\beta$, which governs the scaling
      of the coupling constant and the potential in Eq.(\ref{write})
      equals the lowest coefficient of the Gell-Mann - Low function
      (normalized here in such a way that for one generation of
      leptons $\beta=\beta_{e}=2/3\pi$).  Both the Standard Model and
      experimental data indicate that $\alpha(\mu)$ rises with the
      mass parameter $\mu$, i.e.  $\beta>0$, see
      Ref.\cite{eidelman-et-al_2004} and references therein; the rise
      presumably continues up to the Grand Unification limit
      \cite{footnote}.
      
      Substituting Eqs.(\ref{pot}),(\ref{write}) into Eq.(\ref{Ups})
      one derives
      \begin{eqnarray}
        \label{estUps}
        \Upsilon(r)&\simeq& 
        Z\alpha^2 \beta/(m^2r^3),\quad
        r\rightarrow 0~,
      \end{eqnarray}
      where the lowest term of the $\alpha$-expansion is retained.  It
      is vital that $\Upsilon(r)$ is positive and large, $\Upsilon(r)
      \gg |U(r)|$. This fact makes the effective potential in
      Eq.(\ref{ugh}) also large and positive when $r\rightarrow 0$
         \begin{eqnarray}
        \label{larpos}
        {\cal U}(r)\simeq -\,H(r)\simeq -\,U(r)\,\Upsilon(r)\simeq
          Z^2\alpha^3\beta/(m^2r^4)~.
      \end{eqnarray}
      Thus the vacuum polarization results in an intense repulsion, in
      contrast with a mild attraction, which shows ${\cal U}(r)$ in
      Eq.(\ref{sati}) for a pure Coulomb case.  When the estimate
      Eq.(\ref{larpos}) is applicable, Eq.(\ref{phi''}) allows an
      analytical solution
      \begin{eqnarray}
        \label{re}
        \varphi(r)\propto 
      r \,\exp\left(\!-\frac{Z
        \alpha\,(\alpha \beta)^{1/2}}{mr} \,\right).~~
      \end{eqnarray}
      It shows that $\varphi(r)$ exponentially decreases at small
      distances.  According to Eqs.(\ref{wv}),(\ref{phi}) the
      functions $v(r),w(r)$, also decrease exponentially here;
      correspondingly, the charge density of the $W$-boson
      Eq.(\ref{expr}) decreases exponentially at the origin as well.
      Thus, an account of the QED vacuum polarization eradicates the
      difficulty related to the infinite charge of a vector boson
      located on the Coulomb center.
      
      For small $Z$ the energy shift of discrete energy levels due to
      the vacuum polarization is found to be small (details to be
      reported elsewhere), which makes the Sommerfeld formula derived
      for the spectrum of vector bosons in
      \cite{corben-schwinger_1940} applicable.
     
      The calculations presented raise two vital qualitative
      questions.  First, why the vacuum polarization pushes the system
      in the right direction, reducing the charge density of the
      $W$-boson near the Coulomb center. A simple answer is related to
      the sign of the charge density produced by the polarization.
      Take, for example a positive Coulomb center, $Z>0$. Then the
      vacuum polarization produces negative charge density, $\rho <0
      $. As a result, the potential energy of the $W^-$ boson acquires
      a positive term $\Upsilon =e \rho/m^2>0$ (the charge of $W^-$ is
      negative, $e<0$). A deeper answer is related to the
      renormalizabity of the Standard Model, which implies that by
      renormalizing relevant physical quantities one is {\it bound} to
      obtain sensible physical results. The relevant quantity in
      question is the charge density of a vector boson. Taking into
      account the vacuum polarization, we effectively renormalize the
      coupling constant, which leads to an acceptable physical result.
      
      Second, why the weak vacuum polarization produces a large
      variation of the charge density of the vector boson. The point
      is that the charge density related to the vacuum polarization
      results in an effective potential barrier ${\cal U}=\eta/r^4$
      for vector particles, which cannot be penetrated for arbitrary
      small $\eta$. The answer can be given in more general terms. In
      the pure Coulomb problem the charge density of a vector boson is
      large, ultimately infinite on the Coulomb center. This density
      can be thought of as a quantity, which measures the reaction of
      the boson to the variation of an external electric potential,
      which therefore is very strong. This explains the strong impact
      of the weak vacuum polarization, see more details in
      \cite{kuchiev_flambaum_2005}.
      
      Previous attempts to formulate the Coulomb problem for vector
      bosons within the framework of the Standard Model have been
      facing a difficulty related to an infinite charge of the boson
      located near an attractive Coulomb center.  This work finds that
      the polarization of the QED vacuum resolves the problem.
      Usually the QED radiative corrections produce only small
      perturbations. It is interesting that in the case discussed the
      radiative correction plays a major, defining role.
      
      This work was supported by the Australian Research Council. One
      of us (V.V.F.) is thankful for the support to the Department of
      Energy, Office of Nuclear Physics, Contract No.
      W-31-109-ENG-38.


\begin{thebibliography}{99}
    
  \bibitem{proca_1936}
    
    A. Proca, Compt.Rend. {\bf 202}, 1490 (1936).

  \bibitem{massey-corben_1939}

    H. F. W. Massey and H. C. Corben, Proc.Camb.Phi.Soc. {\bf 35}, 463
    (1939).

  \bibitem{oppenheimer-snyder-serber_1940}
    
    J. R. Oppenheimer, H. Snyder and R. Serber, Phys.Rev. {\bf 57}, 75
    (1940).

  \bibitem{tamm_1940-1-2} 
    
    I. E. Tamm, Phys. Rev. {\bf 58}, 952 (1940); Doklady USSR Acad of
    Sci {\bf 8-9}, 551 (1940).


  \bibitem{corben-schwinger_1940}
    
    H. C. Corben and J. Schwinger, Phys.Rev. {\bf 58}, 953 (1940).
    

  \bibitem{schwinger_1964}
    
    J. Schwinger, Rev. Mod. Phys. {\bf 36}, 609 (1964).
    
    
  \bibitem{cheng_wu_1972}
    
    H. Cheng and T. T. Wu, Phys.Rev.D {\bf 5}, 3247 (1972).

  \bibitem{huang_1992} 
    
    K. Huang, {\it Quarks, Leptons and Gauge Fields}, 2nd edition,
    World Scientific, 1992.
    
  \bibitem{kuchiev_flambaum_2005}
    
    M. Yu. Kuchiev and V. V. Flambaum. To be posted at hep-th soon
    (2005).


  \bibitem{thooft-veltman-1972}

    G. 't Hooft and M.J.G. Veltman, Nucl.Phys. {\bf B44}, 189 (1972).
% REGULARIZATION AND RENORMALIZATION OF GAUGE FIELDS.



  \bibitem{vijayalakshmi-seetharaman-mathews_1979}
    
    B. Vijayalakshmi, M. Seetharaman, and P.M. Mathews, J.Phys.A {\bf
      12}, 665 (1979).

  \bibitem{pomeransky-khriplovich_1998}
    
    A. A. Pomeransky and I. B. Khriplovich, JETP {\bf 86}, 839 (1998).
    

  \bibitem{pomeransky-se'nkov_1999} 
    
    A. A. Pomeransky and R. A. Sen'kov, Phys. Lett. B {\bf  468}, 251
    (1999).
    

  \bibitem{pomeransky-sen'kov-khriplovich_2000}
    
      A. A. Pomeransky, R. A. Sen'kov and I.B. Khriplovich, Phys.Usp.
    {\bf 43} 1055 (2000); Usp.Fiz.Nauk {\bf 43} 1129 (2000).

  \bibitem{silenko_2004}
    
    A. J. Silenko, Analysis of wave equations for spin 1 particles
    interacting with an electromagnetic field, hep-th/0404074.


  \bibitem{fushchych-nikitin-susloparow_1985} 
    
    W. I. Fushchych, A. G. Nikitin, and W. M. Susloparow.  Nuovo
    Cimento {\bf 87}, 415 (1985).


  \bibitem{fushchych-nikitin_1994} 
    
    W. I. Fushchych and A. G. Nikitin.  {\it Symmetries of Equations of
      Quantum Mechanics}, N.Y. Allerton Press, 1994.
    

  \bibitem{sergheyev_1997}
 
    A. G. Sergheyev,    Ukr. J.  Phys. {\bf 42} 1171 (1997).
%    Proc. 2nd Int. Conference "Symmetry in Nonlinear Mathematical
%    Physics" (Kyiv, July 7-13, 1997), Kyiv, 1997, 2, p.331-335.


  \bibitem{weinberg_2001}
    
    S. Weinberg, {\it The quantum theory of feilds, Volume II, Modern
      applications}, Cambridge, University Press, 2001.
    

%  \bibitem{taylor} 
%    
%    J. C. Taylor et al, {\it Gauge Theories of Weak Interactions
%      (Cambridge Monographs on Mathematical Physics)},
%% P. V.  Landshoff (Series Editor), D. R. Nelson (Series Editor), D. W.
%%    Sciama (Series Editor), S. Weinberg (Series Editor)
%    Cambridge University Press, 1976.

 
  \bibitem{LL4}
    
    V. B. Berestetsky, E. M. Lifshits, and L. P. Pitaevsky, {\it Quantum
      electrodynamics}, PergamonPress, 1982.
    

%  \bibitem{akhiezer}
%        
%    A. I. Akhiezer, and V. B. Berestetskii, {\it Quantum
%      Electrodynamics}. New York: Interscience Publishers, 1965.


% \bibitem{green} 
%    
%    H. S. Green, {\it Matrix mechanics}, P.Noordhoff Ltd-Groninger-The
%    Netherlands, 1965.
    
  \bibitem{eidelman-et-al_2004} 
    
    S. Eidelman et al., Phys. Lett. B 592, 1 (2004)
    
  \bibitem{footnote}Numerically $\beta$ can be estimated from two
    reference points $\alpha(m_\tau)$ and $\alpha(M_Z)$ provided in
    Ref.\cite{eidelman-et-al_2004}; this procedure gives $\beta
    \approx 1.41(1) $, though for $\mu \gg m_Z$ there may exist
    corrections due to heavy particles. The normalization on the mass
    $m_Z$ of the $Z$-boson in Eq.(\ref{write}) means that we presume
    that $ \alpha=\alpha(m_Z)\approx 1/128$.

%    \bibitem{alpha}
%      
%      TOPAZ: I. Levine et al., Phys. Rev. Lett. {\bf 78}, 424 (1997);
%      VENUS: S. Okada et al., Phys. Rev. Lett. {\bf 81}, 2428 (1998);
%      OPAL: G.  Abbiendi et al., Eur. Phys. J. C{\bf 13}, 553 (2000);


 \end{thebibliography}
\end{document}